\documentclass[12pt]{article}
\usepackage{amssymb,amsmath,latexsym,graphicx,amsfonts}
\begin{document}
\title{Radiation Reaction in a Lorentz-Violating Electrodynamics}
\author{Veera Reddy Vippala, Sashideep Gutti and Asrarul Haque}
\date{\today}
\maketitle
\abstract{We study radiation reaction in a Lorentz violating electrodynamics \cite{kost1}. We explore the possible modification whatsoever present in the radiation reaction force experienced by an accelerating charge in the modified Maxwell theory. However it turns out that radiation reaction receives  no change due to Lorentz violation whereas electromagnetic mass manifests anisotropy.}
\section{Introduction}
Lorentz invariance and quantum theory are the pillars of Modern Physics. Lorentz invariance serves as one of the most important defining axioms for the two fundamental theories: General Theory of Relativity and Quantum Field Theory.\\
Physics at high energy might unveil the nonlocal structure of the particles or/and the nonlocal interactions, the fundamental length scale and the nontrivial structure of space-time etc. leading to suspect the validity of Lorentz invariance at these scales. \\
A review on the experimental and observational bounds on the relevant scales for the violation of Lorentz symmetry can be found in \cite{mattingly} and the references therein. \\
Lorentz violation in a controlled manner is introduced via phenomenological theories known as Standard Model Extensions. These correspond to modifications suggested to the standard model Lagrangians which break Lorentz symmetry without breaking other symmetries. Tiny terms are added to the Lagrangian which preserve gauge symmetry, conservation of energy-momentum etc., but violate Lorentz invariance \cite{bech,kosteleky}. The question is does such Lorentz violation induce any modification to radiation reaction associated with an accelerating charge?\\
According to the usual classical electrodynamics, an accelerating charge radiates and therefore it loses energy \cite{rr1,rr2,rr3}. Larmor power accounts
for the rate of loss of energy for a non-relativistic accelerated point charge. Loss of energy
could be conceived to be caused by a damping force operating on the charge produced by
the radiation field. This damping force is known as radiation reaction.\\
Although radiation reaction in the usual Maxwell theory has been studied extensively in the past \cite{rr4,rr5,rr6,rr7,rr8,rr9}, it continues
to be the subject-matter of several recent studies \cite{rr10,rr11,rr12,rr13}.\\
The goal of the present work is to compute the forces experienced by an accelerating charge in the birefringent modified Maxwell theory \cite{shreck} which violates Lorentz invariance but preserves $U(1)$ gauge symmetry. 
The current work is a modest endeavor to explore the possible modification (if any) present in the radiation reaction force experienced by an accelerating charge in the modified Maxwell theory. We perform non-relativistic calculation of radiation reaction on an accelerating charge to capture any effect whatsoever arising due to Lorentz violation. We observe that radiation reaction receives no change in the modified Maxwell theory however electromagnetic mass manifests anisotropy.  This work serves as a preliminary framework to continue further study of radiation reaction in the quantum version of Lorentz violating electrodynamics. 
\section{Lorentz-Violating Electrodynamics}
We consider CPT-even Lorentz-violating electrodynamics studied extensively by Kosteleky et al \cite{kosteleky,bailey}.
 The relevant Lagrangian \cite{kost1} density is given by 
\begin{equation}
\mathcal{L}=-\frac{1}{4}F^{\mu\nu}F_{\mu\nu}-\frac{1}{4}(\kappa_{F})_{k\lambda\mu\nu}F^{k\lambda}F^{\mu\nu}-j^\mu A_\mu
\end{equation}
where 
 $F_{\mu\nu}=(\partial_{\mu}A_{\nu}-\partial_{\nu}{A_{\mu}})$ is the field strength tensor.
All fields are defined on Minkowski spacetime with coordinates $(x^\mu)=(x^0,\vec {x})=(ct,x^1,x^2,x^3)$ and a metric of the type
 $ \eta^{\mu\nu} = diag\left(1,-1,-1,-1\right)$. We shall use natural unit $(c = 1)$ throughout this paper. The first term in Eq. (1) is the standard Maxwell term and 
the second represents Lorentz-violating term. The coefficients $(\kappa_{F})_{k\lambda\mu\nu}$ control the Lorentz violation and are expected to be small. Moreover, the four tensor $(\kappa_{F})_{k\lambda\mu\nu}$ transforms covariantly with respect to observer Lorentz transformations but it is fixed with particle Lorentz transformations \cite{coll}. Hence the second term violates particle Lorentz invariance. The background tensor $(\kappa_{F})_{k\lambda\mu\nu}$ is dimensionless and has the symmetries of the Riemann tensor and non vanishing double trace. So, $(\kappa_{F})_{k\lambda\mu\nu}$ contains 19 independent real components.\\
Variation of the Lagrangian (1) yields inhomogeneous equations of motion
\begin{equation}
\partial^{\alpha}F_{\alpha\mu}+(\kappa_{F})_{\mu\alpha\beta\gamma}\ \partial^{\alpha}F^{\beta\gamma}+j_\mu =0.\\
\end{equation} 
 The equations of motion (2) for Lorentz-violating electrodynamics can be cast in the form of the Maxwell equations in homogeneous anisotropic medium given by \cite{bailey}
\begin{equation}
\vec{\nabla} \times \vec{H}-\partial_{0} \vec{D}=\vec{J}, \hspace{1 in} \vec{\nabla}.\vec{D}=\rho$$
$$\vec{\nabla} \times \vec{E}+\partial_{0} \vec{B}=0, \hspace{1 in} \vec{\nabla}.\vec{B}=0
\end{equation}
provided the fields $\vec{D}$ and $\vec{H}$ are related to $\vec{E}$ and $\vec{B}$ as follows:
$$\begin{bmatrix}
\vec{D}\\ \vec{H}
\end{bmatrix}=\begin{bmatrix}
1+k_{DE}	&	k_{DB}\\
k_{HE}		&	1+k_{HB}
\end{bmatrix} \begin{bmatrix}
\vec{E}\\
\vec{B}
\end{bmatrix},$$
where the $ 3 \times 3 $ matrices $k_{DE}$, $k_{DB}$, $k_{HE}$ and $k_{HB}$  are defined by 
\begin{equation}
(k_{DE})^{jk}= -2 (\kappa_F)^{0j0k} $$
$$(k_{HB})^{jk}= \frac{1}{2} {\epsilon}^{jpq} {\epsilon}^{krs}(\kappa_F)^{pqrs}$$
$$(k_{DB})^{jk}= -(k_{HE})^{kj}=(\kappa_F)^{0jpq}{\epsilon}^{kpq}.
\end{equation}
Birefringent modified Maxwell theory \cite{{shreck}} is studied with one non zero Lorentz-violating coefficient $(\kappa_{F})_{0123}$. All others coefficients, which are not related to $(\kappa_{F})_{0123}$ by symmetry arguments, are set to zero. 
 Let $(\kappa_{F})_{0123}={\varepsilon}$, where ${\varepsilon}$ is a small parameter to be determined by experiment.
In Lorentz-violating electrodynamics, Coulomb gauge $\vec{\nabla}.\vec{A}=0$, temporal gauge $A^{0}=0$ and one of the members of the family of Lorentz gauges $\partial_{\mu} A^{\mu}=0$ choices are inequivalent \cite{kost1}. In this work we derive self force equation of a charged particle under Lorentz gauge with only non zero coefficient $(\kappa_{F})_{0123}$.
\section{Potentials in Modified Maxwell Theory}
The equations of motion (2) in the Lorentz gauge $\partial_{\mu} A^{\mu}=0$ reduce to
\begin{equation}
\square{A_{\mu}}+(\kappa_{F})_{\mu\alpha\beta\gamma} \partial^{\alpha}F^{\beta\gamma}+j_{\mu}=0.
\end{equation}
In fact, Eq.(5) comprises of four set of the following equations for $\mu = 0, 1, 2, 3$:
\begin{eqnarray}
\square{A^0}+2\varepsilon\partial^1(\partial^2 A^3-\partial^3 A^2)=-j^{0}\\
\square{A^1}+2\varepsilon\partial^0(\partial^2 A^3-\partial^3 A^2)=-j^{1}\\
\square{A^2}+2\varepsilon\partial^3(\partial^1 A^0-\partial^0 A^1)=-j^{2}\\
\square{A^3}+2\varepsilon\partial^2(\partial^0 A^1-\partial^1 A^0)=-j^{3}
\end{eqnarray}
Using Fourier transforms for $A^{\mu}$ and $j^{\mu}$,
 $$A^{\mu}(\vec{\bf x},t)=\frac{1}{(2\pi)^4}\iint{\tilde{A}^\mu}(\vec{K},\omega)\exp^{-i(\vec{K}.\vec{\bf x}-\omega t)} d\vec{K}d\omega$$
$$j^{\mu}(\vec{\bf x},t)=\frac{1}{(2\pi)^4}\iint{\tilde{j^{\mu}}}(\vec{K},\omega)\exp^{-i(\vec{K}.\vec{\bf x}-\omega t)} d\vec{K}d\omega$$
Eqs. (6 - 9) take the following forms:
\begin{eqnarray}
(\vec{K}^2-\omega^2)\tilde{\varphi}(\vec{K},\omega)+2\varepsilon(K_xK_z\tilde{A_y}-K_xK_y\tilde{A_z})=-\tilde{\rho}(\vec{K},\omega)\\
(\vec{K}^2-\omega^2)\tilde{A_x}(\vec{K},\omega)+2\varepsilon(\omega K_y\tilde{A_z}-\omega K_z\tilde{A_y})=-\tilde{j_x}(\vec{K},\omega)\\
(\vec{K}^2-\omega^2)\tilde{A_y}(\vec{K},\omega)-2\varepsilon(\omega K_z\tilde{A_x}+K_x K_z\tilde{\phi})=-\tilde{j_y}(\vec{K},\omega)\\
(\vec{K}^2-\omega^2)\tilde{A_z}(\vec{K},\omega)+2\varepsilon(\omega K_y\tilde{A_x}+K_x K_y\tilde{\phi})=-\tilde{j_z}(\vec{K},\omega).
\end{eqnarray}
Solving simultaneously Eqs. (10 - 13) (please see Appendix A), we obtain
\begin{eqnarray}
\tilde{\phi}(\vec{K},\omega)=-\frac{\tilde{\rho}(\vec{K},\omega)}{(\vec{K}^2-\omega^2)}+\frac{2\varepsilon K_x K_z}{(\vec{K}^2-\omega^2)^2}\tilde{j_y}-\frac{2\varepsilon K_x K_y}{(\vec{K}^2-\omega^2)^2}\tilde{j_z}\\
\tilde{A_x}(\vec{K},\omega)=-\frac{\tilde{j_x}(\vec{K},\omega)}{(\vec{K}^2-\omega^2)}-\frac{2\varepsilon \omega K_z}{(\vec{K}^2-\omega^2)^2}\tilde{j_y}+\frac{2\varepsilon \omega K_y}{(\vec{K}^2-\omega^2)^2}\tilde{j_z}\\
\tilde{A_y}(\vec{K},\omega)=-\frac{2\varepsilon K_x K_z}{(\vec{K}^2-\omega^2)^2}\tilde{\rho}(\vec{K},\omega)-\frac{2\varepsilon \omega K_z}{(\vec{K}^2-\omega^2)^2}\tilde{j_x}-\frac{\tilde{j_y}}{(\vec{K}^2-\omega^2)}\\
\tilde{A_z}(\vec{K},\omega)=-\frac{\tilde{j_z}(\vec{K},\omega)}{(\vec{K}^2-\omega^2)}+\frac{2\varepsilon K_x K_y}{(\vec{K}^2-\omega^2)^2}\tilde{\rho}(\vec{K},\omega)+\frac{2\varepsilon \omega K_y}{(\vec{K}^2-\omega^2)^2}{\tilde{j_x}}
\end{eqnarray}
The potentials in the leading order in $(\kappa_{F})_{0123}$ via inverse Fourier transform (please see Appendix A) are obtained as follows:
\begin{equation}
\phi(\vec{\bf x},t)=-\frac{1}{4\pi}\int \frac{[\rho(\vec{\bf x}^\prime,t^\prime)]_\textit{ret}}{\vert\vec{\bf x}-\vec{\bf x}^\prime \vert} d\vec{\bf x}^\prime
\end{equation}
\begin{equation}
A_{i}=-\frac{1}{4\pi}\int \frac{j_i  (\vec{\bf x}^\prime,t^\prime)_{ret}}{R} d\vec{\bf x}^\prime +$$
$$\frac{\varepsilon}{4\pi}(\delta_{ix}\delta_{ly}+\delta_{iy}\delta_{lx})\int \frac{[j_l  (\vec{\bf x}^\prime,t^\prime)]_{ret}}{R} d\vec{\bf x}^\prime +[\frac{\partial[j_l (\vec{\bf x}^\prime,t^\prime)]}{\partial t^\prime}]_\textit{ret} \enspace d\vec{\bf x}^\prime
\end{equation}
where $i = x, y, z$ and $l = x, y$. $ t^\prime = t - R $ is the retarded time with $R=|\vec{\bf x}- {\vec{\bf x}}^\prime|$.
\section{Self Force on a Charged Particle in Modified Maxwell Theory}
We derive self-force thinking of a charged particle as an extended object with a charge distribution consisting of a large number of 
small charge elements.
If an extended charged particle moves with non-uniform velocity, the charge elements comprising such charge distribution, 
begin to exert forces on one another. However, these forces
do not cancel out due to retardation giving rise to a net force known as the self-force. Thus,
the radiating extended charged particle experiences a self-force
which acts on the charge
particle itself.\\
The Lorentz self force in usual Maxwell theory is derived by assuming that the charged particle is rigid, instantaneously at rest and the charge distribution is spherically symmetric \cite{rr1}. In the present work, we shall proceed to derive the self force in modified Maxwell theory associated with an accelerated charged particle based on the same set of assumptions as the usual Maxwell theory.\\
The action of a classical charged particle in modified Maxwell theory is 
\begin{equation}
S= S_0 + S_{int}+S_{\textit{modMax}}
\end{equation}
where $S_0$ is the action for free classical  particle, $S_{int}$ contains interaction term and $S_{modMax}$ contains Maxwell and Modified  Maxwell terms. Now\\
\begin{eqnarray}
S&=& -m \int d\lambda \sqrt{\frac{d x^\mu}{d \lambda}  \frac{d x^\nu}{d \lambda} \eta_{\mu \nu}}-\int d^4 x \, j^{\mu} A_{\mu}\nonumber\\
&-&\int d^4 {x} \left(F_{\mu \nu} F^{\mu \nu}-(\kappa_F)^{\mu \nu \varrho \sigma} F_{\mu \nu} F_{\varrho \sigma}\right)
\end{eqnarray}
The equations of motion of the charged particle for variable $x^{\alpha}$ is
\begin{equation}
m \frac{d^2 x^\mu}{ d \tau ^2}=q {F^\mu }_\alpha \frac{d x^\alpha}{d \tau}.
\end{equation} 
Thus the usual Lorentz force equation holds also in the case of modified Maxwell theory \cite{bailey}.
The self force can be calculated
by the space integrals of electromagnetic fields of an accelerated charge as follows: 
\begin{equation}
\vec F^{self} =\int \rho(\vec{\bf x},t) \left[-\vec{\nabla}  {\phi} - \frac{\partial \vec{A}(\vec{\bf x},t)}{\partial t}\right] d \vec {\bf x}.
\end{equation}
Using Eqs. (18) and (19), we obtain (please see Appendix B)
\begin{eqnarray}
F^{self}_i&=&\frac{4}{3}U\, \dot{v_i} - 2 \varepsilon U (\delta_{ix}\delta_{ly}+\delta_{iy}\delta_{lx})\, \dot{v_l}-\frac{2}{3} q^2 \ddot{v_i}
\nonumber\\
&=&{F}_{i}^{em}+{F}_{i}^{rad}\end{eqnarray}
where ${F}_{i}^{em}$ is given by\\
\begin{equation}
{F}_{i}^{em} =\frac{4}{3}U\, \dot{v_i}- 2 \varepsilon U (\delta_{ix}\delta_{ly}+\delta_{iy}\delta_{lx})\, \dot{v_l}.
\end{equation}
and U is defined as the electromagnetic mass, which arises due to the presence of electromagnetic field. The electromagnetic mass is divergent for the point charge ($R\rightarrow 0^{+}$). To have a sensible theory, these infinities are made to absorb via mass renormalization to obtain the physically observable mass \cite{ahaque}.\\
 We can expand ${F}_{i}^{em}$ as follows:
 \begin{equation}
\begin{bmatrix}
{F}_{x}^{em}\\ {F}_{y}^{em}\\ {F}_{z}^{em}\\
\end{bmatrix}=\begin{bmatrix}
\frac{4}{3}U & -2 \varepsilon U & 0\\ -2 \varepsilon U & \frac{4}{3}U & 0\\ 0 & 0 & \frac{4}{3}U
\end{bmatrix}\begin{bmatrix}
a_{x} \\ a_{y} \\ a_{z}
\end{bmatrix}\\
\implies {F}_{i}^{em}=m_{ij} a_j.
\end{equation}
where,
$${m_{ij}}=\begin{bmatrix}
\frac{4}{3}U & -2 \varepsilon U & 0\\ -2 \varepsilon U & \frac{4}{3}U & 0\\ 0 & 0 & \frac{4}{3}U
\end{bmatrix}.$$
We observe that in the Lorentz-violating electrodynamics, the electromagnetic mass of the charged particle appears as a tensor. 
 However, radiation reaction remains intact and is given by
$${F}_{i}^{rad}= \frac{2}{3} q^2 \ddot{v_i}.$$
\section{Conclusion}
We have shown that an accelerating point charge in the modified Maxwell theory experiences the same radiation reaction as that in the usual Maxwell theory. However it turns out that
electromagnetic mass associated with an accelerating charge in this Lorentz-violating electrodynamics exhibits anisotropy unlike the usual Maxwell theory. It is the Lorentz violation that induces mass anisotropy. 
\subsection*{Appendix A: Calculation of the Modified Potentials $\phi$ and $\vec{A}$}
From the symmetry properties of $(\kappa_{F})_{k \lambda \mu \nu}$ with $(\kappa_{F})_{0123}={\varepsilon}$, we have
\begin{eqnarray}
(\kappa_{F})_{0123}={\varepsilon} \hspace{0.3 in} (\kappa_{F})_{1023}=-\varepsilon \hspace{0.3 in}(\kappa_{F})_{2301}=\varepsilon \hspace{0.3 in} (\kappa_{F})_{2310}=-\varepsilon\nonumber\\
(\kappa_{F})_{3210}=\varepsilon \hspace{0.3 in} (\kappa_{F})_{3201}=-\varepsilon \hspace{0.3 in} (\kappa_{F})_{1032}=\varepsilon \hspace{0.3 in} (\kappa_{F})_{0132}=-\varepsilon.
\end{eqnarray}
Equations of motion for $\mu =0,1,2,3$ are respectively given below:
\begin{equation}
\square{A^0}+2\varepsilon\partial^1(\partial^2 A^3-\partial^3 A^2)=-j^{0}
\end{equation}
\begin{equation}
\square{A^1}+2\varepsilon\partial^0(\partial^2 A^3-\partial^3 A^2)=-j^{1}
\end{equation}
\begin{equation}
\square{A^2}+2\varepsilon\partial^3(\partial^1 A^0-\partial^0 A^1)=-j^{2}
\end{equation}
\begin{equation}
\square{A^3}+2\varepsilon\partial^2(\partial^0 A^1-\partial^1 A^0)=-j^{3}.
\end{equation}
Fourier transforms of 
Eqs. (28), (29), (30) and (31) lead to:
\begin{equation}
(\vec{K}^2-\omega^2)\tilde{\varphi}(\vec{K},\omega)+2\varepsilon(K_xK_z\tilde{A_y}-K_xK_y\tilde{A_z})=-\tilde{\rho}(\vec{K},\omega)
\end{equation}
\begin{equation}
(\vec{K}^2-\omega^2)\tilde{A_x}(\vec{K},\omega)+2\varepsilon(\omega K_y\tilde{A_z}-\omega K_z\tilde{A_y})=-\tilde{j_x}(\vec{K},\omega)
\end{equation}
\begin{equation}
(\vec{K}^2-\omega^2)\tilde{A_y}(\vec{K},\omega)-2\varepsilon(\omega K_z\tilde{A_x}+K_xK_z\tilde{\phi})=-\tilde{j_y}(\vec{K},\omega)
\end{equation}
\begin{equation}
(\vec{K}^2-\omega^2)\tilde{A_z}(\vec{K},\omega)+2\varepsilon(\omega K_y\tilde{A_x}+K_xK_y\tilde{\phi})=-\tilde{j_z}(\vec{K},\omega).
\end{equation}
The above equations could be cast in the following matrix form as:
\begin{equation}
\begin{bmatrix}
(\vec{K}^2-\omega^2)    &      0                          &          2\varepsilon K_xK_z          &     -2\varepsilon K_xK_y \\ 

0                       &  (\vec{K}^2-\omega^2)           &         - 2\varepsilon \omega K_z      &       2\varepsilon \omega K_y\\

-2\varepsilon K_xK_z    &-2\varepsilon\omega K_z          &          (\vec{K}^2-\omega^2)         &         0\\

2\varepsilon K_xK_y     & 2\varepsilon\omega K_y          &                    0                  &          (\vec{K}^2-\omega^2)

\end{bmatrix} \begin{bmatrix}
\tilde{\phi}\\
\tilde{A_x}\\
\tilde{A_y}\\
\tilde{A_z}
\end{bmatrix}=-\begin{bmatrix}
\tilde{\rho}\\
\tilde{j_x}\\
\tilde{j_y}\\
\tilde{j_z}
\end{bmatrix}.
\end{equation}
We can have:
$$\begin{bmatrix}
\tilde{\phi}\\
\tilde{A_x}\\
\tilde{A_y}\\
\tilde{A_z}
\end{bmatrix}=-\begin{bmatrix}
(\vec{K}^2-\omega^2)    &      0                          &          2\varepsilon K_xK_z          &     -2\varepsilon K_xK_y \\ 

0                       &  (\vec{K}^2-\omega^2)           &          -2\varepsilon \omega K_z      &       2\varepsilon \omega K_y\\

-2\varepsilon K_xK_z    &-2\varepsilon\omega K_z          &          (\vec{K}^2-\omega^2)         &         0\\

2\varepsilon K_xK_y     & 2\varepsilon\omega K_y          &                    0                  &          (\vec{K}^2-\omega^2)
\end{bmatrix}^{-1} \begin{bmatrix}
\tilde{\rho}\\
\tilde{j_x}\\
\tilde{j_y}\\
\tilde{j_z}
\end{bmatrix}.$$
Neglecting second and higher order terms in $\varepsilon$, we obtain
$$\begin{bmatrix}
\tilde{\phi}\\
\tilde{A_x}\\
\tilde{A_y}\\
\tilde{A_z}
\end{bmatrix}=-\begin{bmatrix}
\frac{1}{(\vec{K}^2-\omega^2)}       &      0  &   \frac{-2\varepsilon K_xK_z}{(\vec{K}^2-\omega^2)^2}   &     \frac{2\varepsilon K_xK_y}{(\vec{K}^2-\omega^2)^2}\\
0  & \frac{1}{(\vec{K}^2-\omega^2)}  &  \frac{2\varepsilon \omega K_z}{(\vec{K}^2-\omega^2)^2}  &  \frac{-2\varepsilon \omega K_y}{(\vec{K}^2-\omega^2)^2}\\
\frac{2\varepsilon K_xK_z}{(\vec{K}^2-\omega^2)^2} &\frac{2\varepsilon\omega K_z}{(\vec{K}^2-\omega^2)^2}& \frac{1}{(\vec{K}^2-\omega^2)}         &  0\\
\frac{-2\varepsilon K_xK_y}{(\vec{K}^2-\omega^2)^2}     & \frac{-2\varepsilon\omega K_y}{(\vec{K}^2-\omega^2)^2} &  0  &\frac{1}{(\vec{K}^2-\omega^2)} \\                
\end{bmatrix} \begin{bmatrix}
\tilde{\rho}\\
\tilde{j_x}\\
\tilde{j_y}\\
\tilde{j_z}
\end{bmatrix}.$$
The above matrix equation leads to the following equations:
\begin{eqnarray}
\tilde{\phi}(\vec{K},\omega)=-\frac{\tilde{\rho}(\vec{K},\omega)}{(\vec{K}^2-\omega^2)}+\frac{2\varepsilon K_xK_z}{(\vec{K}^2-\omega^2)^2}\tilde{j_y}-\frac{2\varepsilon K_x K_y}{(\vec{K}^2-\omega^2)^2}\tilde{j_z}\\
\tilde{A_x}(\vec{K},\omega)=-\frac{\tilde{j_x}(\vec{K},\omega)}{(\vec{K}^2-\omega^2)}-\frac{2\varepsilon \omega K_z}{(\vec{K}^2-\omega^2)^2}\tilde{j_y}+\frac{2\varepsilon \omega K_y}{(\vec{K}^2-\omega^2)^2}\tilde{j_z}\\
\tilde{A_y}(\vec{K},\omega)=-\frac{2\varepsilon K_x K_z}{(\vec{K}^2-\omega^2)^2}\tilde{\rho}(\vec{K},\omega)-\frac{2\varepsilon \omega K_z}{(\vec{K}^2-\omega^2)^2}\tilde{j_x}-\frac{\tilde{j_y}}{(\vec{K}^2-\omega^2)}\\
\tilde{A_z}(\vec{K},\omega)=-\frac{\tilde{j_z}(\vec{K},\omega)}{(\vec{K}^2-\omega^2)}+\frac{2\varepsilon K_x K_y}{(\vec{K}^2-\omega^2)^2}\tilde{\rho}(\vec{K},\omega)+\frac{2\varepsilon \omega K_y}{(\vec{K}^2-\omega^2)^2}{\tilde{j_x}}.
\end{eqnarray}
Inverse Fourier transforms of the above equations yield:
 \begin{equation}
\phi(\vec{\bf x},t)= - \frac{1}{4\pi}\int\frac{[\rho(\vec{\bf x}^\prime,t^\prime)]_\textit{ret}}{|\vec{\bf x}-\vec{\bf x}^\prime|}d\vec{\bf x}^\prime
 \end{equation}
\begin{equation}
A_x(\vec{\bf x},t)= - \frac{1}{4\pi}\int\frac{\left[j_x(\vec{\bf x}^\prime,t^\prime)\right]_\textit{ret}}{\vert\vec{\bf x}-\vec{\bf x}^\prime\vert}d\vec{\bf x}^\prime$$
$$ + \frac{\varepsilon}{4\pi}\int \left[\frac{\partial{j_y(\vec{\bf x}^\prime,t^\prime)}}{\partial t^\prime}\right]_\textit{ret} \enspace d\vec{\bf x}^\prime+\frac{\varepsilon}{4\pi}\int \frac{\left[j_y(\vec{\bf x}^\prime,t^\prime)\right]_\textit{ret}}{\vert \vec{\bf x}-\vec{\bf x}^\prime\vert}\enspace d\vec{\bf x}^\prime 
\end{equation}
\begin{equation}
A_y(\vec{\bf x},t)= - \frac{1}{4\pi}\int\frac{\left[j_y(\vec{\bf x}^\prime,t^\prime)\right]_\textit{ret}}{\vert \vec{\bf x}-\vec{\bf x}^\prime \vert}d\vec{\bf x}^\prime$$
$$ + \frac{\varepsilon}{4\pi}\int \left[\frac{\partial{j_x(\vec{\bf x}^\prime,t^\prime)}}{\partial t^\prime}\right]_\textit{ret} 
\enspace d\vec{\bf x}^\prime+\frac{\varepsilon}{4\pi}\int \frac{\left[j_x(\vec{\bf x}^\prime,t^\prime)\right]_\textit{ret}}
{\vert \vec{\bf x}-\vec{\bf x}^\prime \vert}\enspace d\vec{\bf x}^\prime
\end{equation}
\begin{equation}
A_z(\vec{\bf x},t)= - \frac{1}{4\pi}\int\frac{\left[j_z(\vec{\bf x}^\prime,t^\prime)\right]_\textit{ret}}{\vert\vec{\bf x}-
\vec{\bf x}^\prime \vert}d\vec{\bf x}^\prime
\end{equation}
Integration over azimuthal angle associated with the terms containing $K_x$ and $K_y$ vanish after inverse Fourier transforms.
Eqs. (42), (43) and (44) could be compactly expressed as:
\begin{eqnarray}
A_{i}&=& - \frac{1}{4\pi}\int \frac{j_i  (\vec{\bf x}^\prime,t^\prime)_{ret}}{R} d\vec{\bf x}^\prime\nonumber\\
&+& \frac{\varepsilon}{4\pi}(\delta_{ix}\delta_{ly}+\delta_{iy}\delta_{lx})\int \frac{[j_l  (\vec{\bf x}^\prime,t^\prime)]_{ret}}{R} d\vec{\bf x}^\prime+[\frac{\partial[j_l (\vec{\bf x}^\prime,t^\prime)]}{\partial t^\prime}]_\textit{ret} \enspace d\vec{\bf x}^\prime \nonumber \\
&=&{A}_{ui}+ {A}_{ci}
\end{eqnarray}
where,
\begin{equation}
A_{ui}=-\frac{1}{4\pi}\int \frac{j_i  (\vec{\bf x}^\prime,t^\prime)_{ret}}{R} d\vec{\bf x}^\prime
\end{equation}
\begin{equation}
A_{ci}= \frac{\varepsilon}{4\pi}(\delta_{ix}\delta_{ly}+\delta_{iy}\delta_{lx})\int \frac{[j_l  (\vec{\bf x}^\prime,t^\prime)]_{ret}}{R} d\vec{\bf x}^\prime+[\frac{\partial[j_l (\vec{\bf x}^\prime,t^\prime)]}{\partial t^\prime}]_\textit{ret} \enspace d\vec{\bf x}^\prime.
\end{equation}
\subsection*{Appendix  B: Calculation of Self Force}
Self force on a rigid charged particle reads
\begin{eqnarray}
\vec{F}_{self}& = &-\int\rho(\vec{\bf x},t)\enspace\left[\triangledown \phi(\vec{\bf x},t) +\frac{\partial\vec{A}(\vec{\bf x},t)}{\partial t} \right]\, d\vec{\bf x}\\
&=& \vec{F}_{u}+ \vec{F}_{c}\nonumber \end{eqnarray}
where $\vec{F}_{u}$ stands for the uncorrected self force and $\vec{F}_{c}$ is corrected part of the self force due to the modified Maxwell's equations. We have
$$\vec{F}_{u}=-\int\rho(\vec{\bf x},t)\enspace\left[\triangledown \phi(\vec{\bf x},t) +\frac{\partial \vec{A}_{u}(\vec{\bf x},t)}{\partial t} \right]\, d\vec{\bf x}$$
The uncorrected piece of self force \cite{rr1} is given by
$$F_{ui}=\frac{4}{3}U\, \dot{v_i}-\frac{2}{3} q^2 \ddot{v_i}$$ 
In order to calculate $\vec{F}_{c}$, we shall expand retarded quantity in a Taylor series as follows,
\begin{equation}
\left[...\right]_{ret} = \sum_{n=0}^{\infty}\frac{(-1)^n}{n!} R^n \frac{\partial^n}{\partial t^n}\left[...\right]_{t^\prime =t}.
\end{equation}
Corrected piece of the self force now reads
\begin{equation}
{F}_{ci}=-\frac{\varepsilon}{4\pi}(\delta_{ix}\delta_{ly}+\delta_{iy}\delta_{lx})\left\lbrace\int \rho(\vec{\bf x},t) d\vec{\bf x}\quad \sum_{n=0}^{\infty} \frac{(-1)^n}{n!}\frac{\partial^{n+1}}{\partial t^{n+1}}\int R^{n-1} \rho(\vec{\bf x^\prime}) v_l(t) d\vec{\bf x}^\prime + \right.$$
$$\left.\int \rho(\vec{\bf x},t) d\vec{\bf x} \quad \sum_{n=0}^{\infty}\frac{(-1)^n}{n!}\frac{\partial^{n+2}}{\partial t^{n+2}}\int R^n \rho(\vec{\bf x^\prime}) v_l(t) d\vec{\bf x}^\prime \right\rbrace
\end{equation}
And
\begin{equation}
F_{ci}=-\frac{\varepsilon}{4\pi}(\delta_{ix}\delta_{ly}+\delta_{iy}\delta_{lx})\left\lbrace \frac{\partial v_l}{\partial t} \iint \frac{\rho(\vec{\bf x},t)\,\rho({\vec{\bf x}^\prime})}{R} d\vec{\bf x} d{\vec{\bf x}}^\prime - \right.$$
$$\left.\frac{\partial^2 v_l}{\partial t^2} \iint \rho(\vec{\bf x},t)\, \rho({\vec{\bf x}^\prime}) \, d\vec{\bf x}\, d{\vec{\bf x}^\prime}\right.+$$
$$\left.\frac{\partial^2 v_l}{\partial t^2} \iint \rho(\vec{\bf x},t)\, \rho({\vec{\bf x}^\prime}) \, d\vec{\bf x}\, d{\vec{\bf x}^\prime}-\right.$$
$$\left.\frac{1}{2}\frac{\partial^3 v_l}{\partial t^3} \iint  R\, \rho(\vec{\bf x},t)\, \rho({\vec{\bf x}^\prime}) d\vec{\bf x} \,d{\vec{\bf x}^\prime}
 + .....\right\rbrace  
\end{equation}
In the point charge limit, the above expression becomes 
\begin{eqnarray} {F}_{ci}&=&-\frac{\varepsilon}{4\pi}(\delta_{ix}\delta_{ly}+\delta_{iy}\delta_{lx}) \frac{\partial v_l}{\partial t} \iint \frac{\rho(\vec{\bf x},t)\,\rho({\vec{\bf x}^\prime})}{R} d\vec{\bf x} d{\vec{\bf x}}^\prime\nonumber\\
&=& - 2 \varepsilon U (\delta_{ix}\delta_{ly}+\delta_{iy}\delta_{lx}) \frac{\partial v_l}{\partial t}.\end{eqnarray}
Thus the self force on a charged particle in the modified Maxwell theory is
\begin{equation}
F^{self}_{i}=\frac{4}{3}U\, \dot{v_i}- 2 \varepsilon U (\delta_{ix}\delta_{ly}+\delta_{iy}\delta_{lx})\, \dot{v_l}-\frac{2}{3} q^2 \ddot{v_i}.
\end{equation}

\end{document}